\documentclass[amsmath,amssymb,ams,superscriptaddress,footinbib,floatfix, citeautoscript]{revtex4-2}

\usepackage[a4paper,textwidth=18.3cm,textheight=26cm]{geometry}
\usepackage[english]{babel}
\usepackage{ulem}

\usepackage{svg}
\usepackage{upgreek}
\usepackage[mathlines]{lineno}
\usepackage{graphicx}
 
\usepackage{xcolor}
\usepackage[T1]{fontenc}
\usepackage{helvet}
\usepackage[version=4]{mhchem}
\usepackage{wasysym}
\usepackage{amssymb}

\usepackage{array}
\newcolumntype{P}[1]{>{\centering\arraybackslash}p{#1}}

\usepackage{xcolor,colortbl}

\definecolor{gold}{HTML}{ffe484}

\newcolumntype{G}[1]{>{\columncolor{gold}\centering\arraybackslash}p{#1}}

\usepackage{fourier} 
\usepackage{array}
\usepackage{makecell}

\usepackage[many]{tcolorbox}
\newtcolorbox{cross}{blank,breakable,parbox=False,
  overlay={\begin{scope}, \draw[red,line width=5pt] (interior.south west)--(interior.north east);
    \draw[red,line width=5pt] (interior.north west)--(interior.south east);\end{scope}}}

\addto\captionsenglish{\renewcommand{\figurename}{Fig.}}

\renewcommand*{\thesection}{\arabic{section}}

\begin{document}

\setcitestyle{super}
\title{Coherent control of a superconducting qubit using light}

\begin{abstract}
    Quantum communications technologies require a network of quantum processors connected with low loss and low noise communication channels capable of distributing entangled states. Superconducting microwave qubits operating in cryogenic environments have emerged as promising candidates for quantum processor nodes. However, scaling these systems is challenging because they require bulky microwave components with high thermal loads that can quickly overwhelm the cooling power of a dilution refrigerator. Telecommunication frequency optical signals, meanwhile, can be fabricated in significantly smaller form factors while avoiding challenges due to high signal loss, noise sensitivity, and thermal loads due to their high carrier frequency and propagation in silica optical fibers. Transduction of information via coherent links between optical and microwave frequencies is therefore critical to leverage the advantages of optics for superconducting microwave qubits, while also enabling superconducting processors to be linked with low-loss optical interconnects. Here, we demonstrate coherent optical control of a superconducting qubit. We achieve this by developing a microwave-optical quantum transducer that operates with up to 1.18\% conversion efficiency with low added microwave noise, and demonstrate optically-driven Rabi oscillations in a superconducting qubit.  
    \end{abstract}

\author{Hana K. Warner}
 \affiliation{Harvard John A. Paulson School for Engineering and Applied Sciences, Cambridge, MA 02138, USA}

\author{Jeffrey Holzgrafe}
\altaffiliation[Now at ]{Hyperlight Corporation, Cambridge, MA, 02138, USA}
 \affiliation{Harvard John A. Paulson School for Engineering and Applied Sciences, Cambridge, MA 02138, USA}

\author{Beatriz Yankelevich}
\altaffiliation[Now at ]{Research Laboratory of Electronics, Massachusetts Institute of Technology, Cambridge, MA 02139, USA}
\affiliation{Rigetti Computing, 775 Heinz Avenue, Berkeley, CA 94710, USA} 

\author{David Barton}
\altaffiliation[Now at ]{Department of Materials Science and Engineering, Northwestern University, Evanston, Illinois 60208, USA}
 \affiliation{Harvard John A. Paulson School for Engineering and Applied Sciences, Cambridge, MA 02138, USA}

\author{Stefano Poletto}
\affiliation{Rigetti Computing, 775 Heinz Avenue, Berkeley, CA 94710, USA}

\author{C. J. Xin}
 \affiliation{Harvard John A. Paulson School for Engineering and Applied Sciences, Cambridge, MA 02138, USA}

\author{Neil Sinclair}

 \affiliation{Harvard John A. Paulson School for Engineering and Applied Sciences, Cambridge, MA 02138, USA}
 \affiliation{Division of Physics, Mathematics and Astronomy, and Alliance for Quantum Technologies (AQT), California Institute of
Technology, Pasadena, California 91125, USA}

\author{Di Zhu}
 \affiliation{Harvard John A. Paulson School for Engineering and Applied Sciences, Cambridge, MA 02138, USA}

\author{Eyob Sete}
\affiliation{Rigetti Computing, 775 Heinz Avenue, Berkeley, CA 94710, USA}

\author{Brandon Langley}
\affiliation{Rigetti Computing, 775 Heinz Avenue, Berkeley, CA 94710, USA}

\author{Emma Batson}
\affiliation{Research Laboratory of Electronics, Massachusetts Institute of Technology, Cambridge, MA 02139, USA}

\author{Marco Colangelo}
\altaffiliation[Now at ]{Electrical and Computer Engineering Department, Northeastern University, 360 Huntington Avenue, Boston, MA 02115, USA}
\affiliation{Research Laboratory of Electronics, Massachusetts Institute of Technology, Cambridge, MA 02139, USA}

\author{Amirhassan Shams-Ansari}
 \affiliation{Harvard John A. Paulson School for Engineering and Applied Sciences, Cambridge, MA 02138, USA}

\author{Graham Joe}
 \affiliation{Harvard John A. Paulson School for Engineering and Applied Sciences, Cambridge, MA 02138, USA}

\author{Karl K. Berggren}
\affiliation{Research Laboratory of Electronics, Massachusetts Institute of Technology, Cambridge, MA 02139, USA}

\author{Liang Jiang}
\affiliation{Pritzker School of Molecular Engineering, University of Chicago, Chicago, IL 60637, USA}

\author{Matthew Reagor}
\affiliation{Rigetti Computing, 775 Heinz Avenue, Berkeley, CA 94710, USA}

\author{Marko Lon\v{c}ar}
 \affiliation{Harvard John A. Paulson School for Engineering and Applied Sciences, Cambridge, MA 02138, USA}
\maketitle

\subsection{Introduction} 
Superconducting (SC) qubits and SC quantum circuits have emerged as one of the most promising quantum computing (QC) platforms and have been used to demonstrate processing advantages over classical supercomputers, even in the presence of system noise \cite{arute.19, kim.23}. However, to reach the true potential of this platform, systems containing hundreds of logical qubits (many thousands of physical qubits) are required \cite{gambetta.17}. This is challenging since SC qubits require ultra-low temperatures to operate, and this large number of qubits would result in prohibitively large dilution refrigerators with cooling power that cannot be achieved with current technology \cite{krinner.19}. 
One approach to overcome this relies on a modular computing scheme \cite{krastanov.21, zhong.22} based on a network of small scale quantum processors, each in its own refrigerator, connected with low noise and low loss quantum links. Since SC qubits can be accessed using microwave photons with frequency $\sim$3-8 GHz range, SC quantum links between dilution refrigerators could be used to allow for the transmission of microwave  signals. However this requires cryogenically cooled (<50 mK) links between each subsystem \cite{storz.23}, which is challenging, costly, and not scalable.

\begin{figure*}[!htb]
    \centering
    \includegraphics{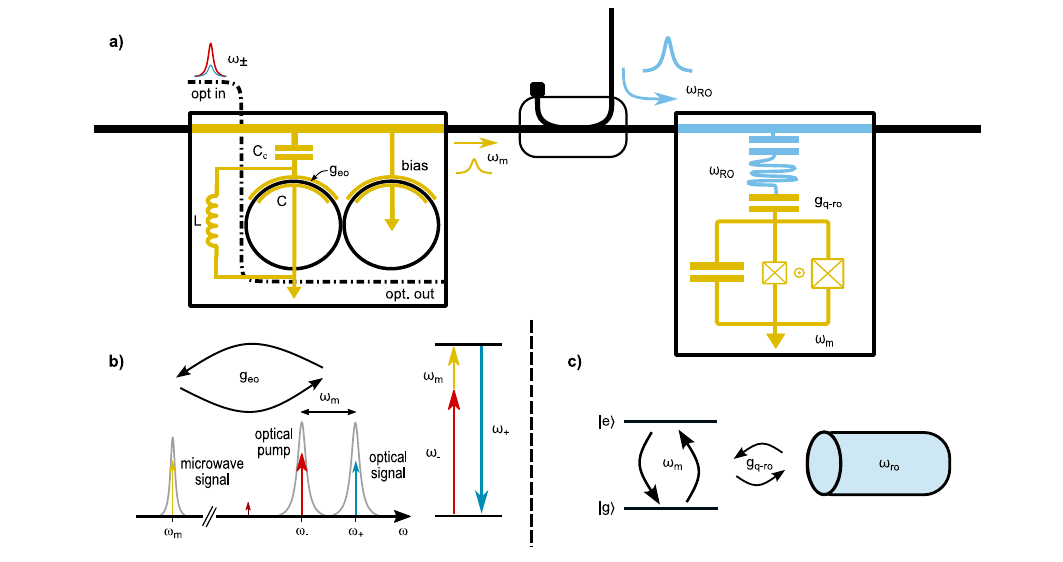}
    \caption{Transducer-driven superconducting qubit scheme. (a) Two optical fields ($\omega_\pm$) are coupled into the transducer via a waveguide. The two modes are resonant with the transducer's hybridized optical modes which are generated by a coupled paperclip resonator capacitively coupled ($C_C$) to a microwave LC resonator ($\omega_m=\frac{1}{\sqrt{LC}}$). The optical modes interact in the transducer to generate a microwave tone ($\omega_m$) via difference frequency generation using the $\chi^{(2)}$ nonlinearity in TFLN according to microwave-optical coupling rate $g_{eo}$. The microwave tone is then transmitted to a SC qubit chip via a coaxial cable to drive the qubit through a dispersively coupled readout resonator. We measure the qubit state by measuring the transmission of a readout pulse ($\omega_{ro}$) generated at room temperature. (b) Schematic of cavity electro-optic transduction process. Two hybridized optical modes ($\omega_\pm$) are detuned by the microwave resonance frequency ($\omega_m$).  Energy exchange between the microwave and optical domains are mediated by the strong pump field at $\omega_-$.  (c) The qubit computational ground ($|g\rangle$) and excited ($|e\rangle$) states are separated by frequency $\omega_m$.  The qubit is dispersively coupled to a microwave readout resonator ($\omega_{ro}$) with strength $g_{q-ro}$. }
    \label{fig:schem}
\end{figure*}

An alternative approach relies on interfacing SC qubits with photons at telecommunication frequencies ($\sim$200 THz), propagating in low-loss optical fibers. Optical photons have much higher energy than microwave photons, and thus are insensitive to thermal noise even at room temperature. Furthermore, their high carrier frequency and large bandwidth allow for wavelength-division multiplexing, resulting in a 100-fold increase in the number of addressable physical qubits in a single fiber. Moreover, since silica optical fibers are weak carriers of thermal energy, they could be used to replace  traditional microwave coaxial cables \cite{lecocq.21} to provide a 1000x reduction in thermal load for signals routed  in and out of a refrigerator. Finally, optical fiber-based quantum links between SC processors benefit from mature classical optical communication technology and could enable efficient, low-noise, and high rate quantum interconnects between quantum processors over kilometers \cite{bersin.23}. This reach can be further extended by combining processors with emerging quantum-repeater based quantum networks \cite{reiserer.14}.  

A key component needed to interface SC qubits with light is a microwave-optical quantum transducer (MOQT), a device capable of providing a coherent, bidirectional link while preserving fragile quantum states across five orders of magnitude of energy. This is challenging due to the lack of a strong coherent nonlinearity that can bridge  this vast energy gap between microwave and optical photons. Microwave-optical quantum transduction has been pursued using several approaches \cite{lauk.20, clerk.20, han.21}.  Piezoelectric opto-mechanical \cite{jiang.20, weaver.22, mirhosseini.20}, membrane optomechanical \cite{brubaker.22} and cavity electro-optic \cite{sahu.23,xu.19,holzgrafe.20,mckenna.20} platforms have showed particular promise in recent years due to demonstrations of low noise performance \cite{hease.20,Xu.21,brubaker.22}, single photon operation \cite{meesala.23,jiang.23,sahu.23}, and bidirectional conversion \cite{jiang.20,hease.20,brubaker.22,fan.18,weaver.22,andrews.14}. Furthermore, recent demonstrations of optical driving \cite{lecocq.21} and readout of SC qubits \cite{lecocq.21, delaney.22, van_thiel.23, arnold.23} demonstrate the potential to reduce passive heat load and space requirements to enable further scaling of SC processors, with high fidelity SC qubit readout in piezoelectric opto-mechanical \cite{van_thiel.23} and all-optical SC qubit readout in bulk electro-optic \cite{arnold.23}  platforms illustrating the potential for near-term scaling of SC qubit systems with MOQTs. 

Cavity electro-optic (CEO) MOQTs, that leverage strong Pockel's effect in materials with large  $\chi^2$ nonlinearities, are especially of interest due to the direct conversion process between the microwave and optical domains, which circumvents potential lossy,  noisy, or rate-limiting intermediary modes that can limit transduction bandwidth. Among these, the thin-film lithium niobate (TFLN) photonic platform has emerged as a front runner \cite{holzgrafe.20, mckenna.20, han.21}  since it combines a large electro-optic (EO) coefficient ($r_{33}\approx 30$ pm/V), the ability to realize ultra-high Q optical resonators \cite{zhang.17, shams-ansari.22}, and wafer-scale fabrication processes \cite{luke.20}.

Despite great progress, CEO-MOQTs still require strong pump powers which can lead to increased noise and limit the device repetition rates (in order to maintain low bath temperature) \cite{Xu.21, sahu.23}, with open questions on the influence of scattered optical light and DC flux noise on the performance of SC qubits. 

Here, we overcome these limitations and demonstrate a TFLN CEO-MOQT design and use it to demonstrate coherent optical driving of a SC qubit (Fig. \ref{fig:schem}a). The transducer is a triply-resonant system \cite{soltani.17} that consists of an on-chip microwave LC resonator operating at frequency $\omega_m$ that is capacitively coupled via the EO effect to two coupled optical resonators that give rise to hybridized optical modes at frequencies $\omega_+$ and $\omega_- = \omega_+-\omega_m$ \cite{zhang.19}. This allows for a resonantly-enhanced exchange of energy between the microwave and optical modes. In our approach, an optical pump ($\omega_-$) and idler ($\omega_+$) interact at a CEO-MOQT to generate a microwave signal at the qubit frequency ($\omega_m$) via a difference frequency generation process (Fig. \ref{fig:schem}b).

The microwave pulse is then transmitted via a coaxial cable to a qubit in the same refrigerator ($T_{CEO-MOQT}\approx T_{qubit}\approx 15$ mK). The qubit is dispersively coupled to a readout resonator ($\omega_{ro}$), which allows the  read-out of the qubit state via a quantum non-demolition measurement, by measuring the transmission of a readout pulse \cite{krantz.19} (Fig. \ref{fig:schem}c). 
The co-operation of a thin-film CEO-MOQT and SC qubit allows us to characterize the impact of the thin-film CEO-MOQT operation on the SC qubit and validate near-term utility of TFLN electro-optic transducers for optically interfacing with SC qubits. 
\subsection{Electro-optic transducer characterization}
\begin{figure*}[!htb] 
    \centering
    \includegraphics{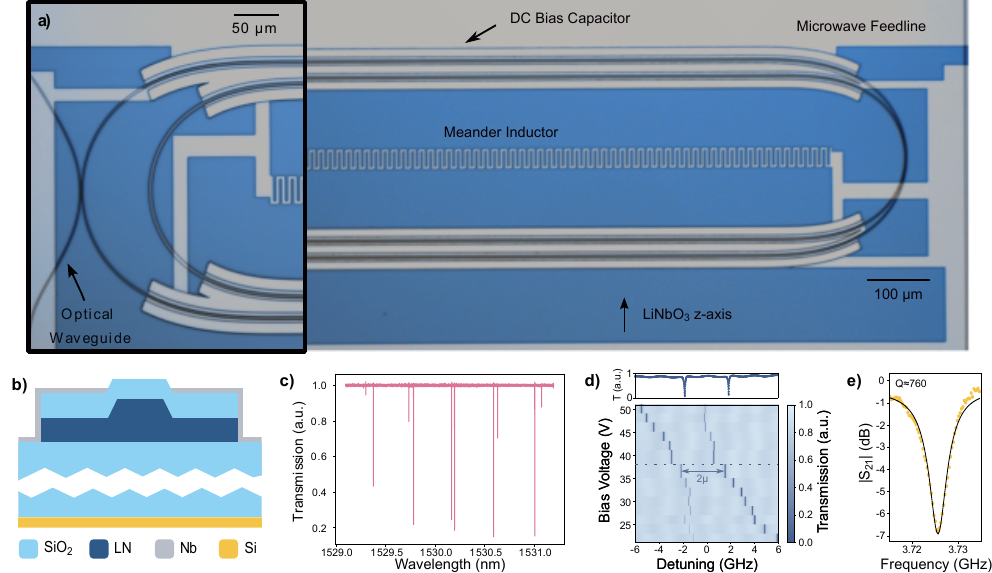}
    \caption{Cavity electro-optic microwave-optic quantum transducer (CEO-MOQT). (a) Optical micrograph of transducer. A niobium (Nb) microwave LC resonator (silver) is capacitively coupled to two hybridized lithium niobate racetrack resonators in a paperclip geometry (black) to coherently exchange energy between the microwave and optical domains using the resonantly-enhanced electro-optic effect in lithium niobate. This geometry allows us to have a small device footprint to lower parasitic capacitance. The device is cladded with silica to aid with thermal dissipation and mitigate optical losses due to electrical crossovers. The microwave signal is read out via a transmission line (top), and the optical signal is collected via optical grating couplers (See Extended Data Fig. \ref{fig:gcs}). The optical racetrack resonators adiabatically taper from 1 $\upmu$m in coupling regions to 3 $\upmu$m in the straight section of the racetrack to reduce sidewall scattering losses while supporting only the fundamental transverse mode (inset). (b)  Drawing of device cross-section. Our optical waveguides are gated in a plateau-like electrode geometry to directly contact our Nb electrodes with the z-axis of the LN crystal, thus benefiting from the large Pockel's coefficient ($r_{33}\approx$ 30 pm/V). This improves microwave-optical field overlap, and thus the single photon coupling rate between our microwave and optical modes. (c) The nested paperclip geometry results in a Vernier effect with 5 nm periodicity and ensures the existence of a pair of hybridized optical modes with splitting similar to the microwave resonant frequency. (d) We can then ensure that we operate at the triple-resonance point by controlling the optical detuning with a voltage bias ($2\mu\approx$ 3.5 GHz) in order to match (e) our microwave resonator frequency ($\omega_m\approx$ 3.7 GHz).}
    \label{fig:ceo}
\end{figure*}
Our transducer is realized using coupled optical racetrack resonators in a paperclip configuration (Fig. \ref{fig:ceo}a) that supports hybridized optical modes that are delocalized between the two racetrack resonators and feature a characteristic split resonance. 

The paperclip resonator allows us to maintain a low parasitic capacitance due to the small device footprint. Our optical resonators feature a 1 $\upmu$m wide optical waveguide in coupling regions in order to prevent exciting higher order transverse modes. The waveguide is then adiabatically expanded to 3 $\upmu$m in straight sections of the cavities to minimize optical losses caused by sidewall roughness ($\kappa_{o,i}/2\pi\approx 25$ MHz, $\kappa_o/2\pi\approx 80$ MHz). Waveguides are patterned using a negative tone e-beam resist (hydrogen silsesquioxane FOx-16) and etched using Ar+ plasma ion etching. After etching, devices are annealed in a 520$^{\circ}$C oxygen environment to mitigate defects in as-received lithium niobate or accumulated during the etching process. They are then cladded with IC-PECVD silicon dioxide and re-annealed at 520$^\circ$C in $O_2$ to mitigate hydrogen defects in the device oxide cladding that result in excess optical losses \cite{shams-ansari.22}. Next, windows in our oxide layer surrounding optical waveguides are created using a combination of \ce{Ar+} and \ce{CF4} reactive ion etching. Nb is then sputtered and patterned using negative tone photoresist (SPR 700-1.0) and etched using \ce{CF4} reactive ion etching to create superconducting electrodes needed to realize our microwave LC resonator and transmission lines.  This allows us to use a plateau-like electrode geometry (Fig. \ref{fig:ceo}b), where our electrodes are in direct contact with the z-axis of the \ce{LiNbO3} crystal to enhance the microwave-optical field overlap. This improves our simulated microwave-optical coupling rate by 40\% when compared to a fully cladded device structure and increases the rate of coherent exchange of energy between optical and microwave signals (see Extended Data Fig. \ref{fig:g0}). Furthermore, direct contact between LN and metal is known to mitigate charge screening effects and DC bias drift \cite{holzgrafe.24}, enabling a more stable operation at the triple-resonance condition.

A 191 $\upmu$m mismatch in the length of the coupled racetrack resonators is introduced to create a 4 GHz (31 pm) difference in free-spectral range, which results in a Vernier effect for our optical modes (Fig. \ref{fig:ceo}c) with 5 nm periodicity. This results in different coupling and frequency splitting (anti-crossing) for different modes and ensures that some of the hybridized mode pairs feature a resonance split close to the microwave resonator frequency. Then, moderate DC bias voltages ($|V_{DC}|<50$ V) compatible with our cryogenic bias-tee at the mixing chamber plate of the refrigerator can be applied to bias the electrode to fine-tune the resonance splitting via the EO effect and achieve our triple resonance condition (Fig. \ref{fig:ceo}d).

Our LC microwave resonator is designed to be a few MHz blue-detuned from the desired microwave frequency $\omega_m=\omega_+-\omega_-$ to compensate for red-shifting and broadening it experiences under strong optical pumping due to small but finite optical absorption in our superconductor. The device reported here has a microwave frequency $\omega_m/2\pi\leq 3.71$ GHz (Fig \ref{fig:ceo}e) and a minimal optical mode splitting $2\mu\approx 3.5$ GHz (Fig \ref{fig:ceo}d). We operated our device with a DC voltage bias $V_{DC}\approx36$ V.  

To  characterize the efficiency of the bidirectional transduction process ($\eta$), we excite the paperclip resonator with a strong optical pump resonant with our red optical mode ($\omega_-\approx 193$ THz). For optical-to-microwave conversion, we also excite the paperclip resonator with a weaker optical idler resonant with the blue optical mode ($\omega_+$), that is derived from the pump  using a single sideband modulator (SSBM).  The idler power is 20 dB lower than the pump power in order to remain in the red-pumping regime. The two modes interact in the transducer to generate a microwave signal at the microwave resonator frequency via difference frequency generation ($\omega_m=\omega_+-\omega_-$), which we measure on a network analyzer after $\sim$ 70 dB of link gain. In the microwave-optical conversion regime, a microwave tone generated by the network analyzer interacts with the red pump to generate a blue optical signal via sum frequency generation ($\omega_+=\omega_-+\omega_m$).  The on-chip microwave power is kept low, with uniform power $P_m\approx-80$ dBm.  We measure the beatnote of the optical signal against an optical pump on a calibrated InGaAs detector in order to extract the flux of the optical signal and characterize conversion in the microwave-optical regime (Methods \ref{meth:eff}). The device is characterized in the continuous wave (CW) and pulsed optical regimes. The pulsed regime allows for larger peak optical powers to be applied while avoiding negative photorefractive effects or thermal bath heating. Square optical pulses are generated by modulating the state of our optical fields by an acousto-optic modulator (AOM) driven by an arbitrary-waveform generator (AWG). Pulses are then amplified by an erbium-doped fiber amplifier (EDFA) before being sent to the transducer. In the pulsed-optical regime, we choose our optical pulse width and duty cycle to maintain a low thermal bath temperature ($T_{bath}\leq 30$ mK). See Extended Data Fig. \ref{fig:full_schem} for further details on the measurement system. 

\begin{figure}[!htb]
    \centering
    \includegraphics{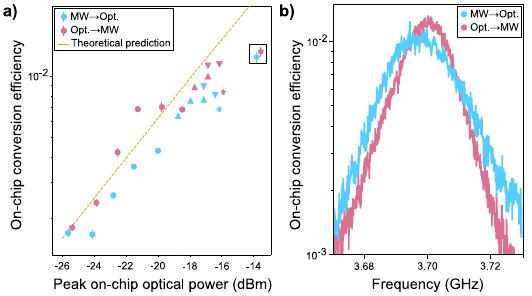}
   \caption{Characterization of transducer performance. (a) On-chip conversion efficiency for MW$\rightarrow$Opt. (blue, sum frequency generation) and Opt.$\rightarrow$MW (pink, difference frequency generation) measured for different peak on chip optical pump powers. The roll-off in conversion efficiency is attributed to optical-pump induced quasi-particle generation in our superconductor resulting in excess loss in our microwave resonator and drift away from the triply resonant condition. $\circ$ markers correspond to CW optical signals, whereas other shapes correspond to a 150 ns pulse with $\triangle$ 1 MHz repetition rate (15\% duty cycle), $\triangledown$ 500 kHz repetition rate (7.5\% duty cycle), $\star$ 333 kHz repetition rate (5\% duty cycle), and $\pentagon$ 200 kHz repetition rate (2\% duty cycle). The theoretical peak conversion efficiency assuming uniform optical and microwave loss rates is plotted as a dotted gold line. (b) A frequency-dependent efficiency sweep for a 150 ns pulse with 200 kHz duty cycle is highlighted in the black box in (a). }
    \label{fig:perf}
\end{figure}

In Fig. \ref{fig:perf}a, we show the on-chip conversion efficiency for different peak on-chip optical powers and duty cycles in the range 2-15\%. For pulsed measurements, we maintained a pulse width of 150 ns with pulse frequency changing from 1 MHz to 200 kHz. Smaller duty cycle is beneficial for keeping average optical power in the refrigerator low, while a high repetition rate can result in faster operation of the transducer. We extract the conversion efficiency per-microwatt of pump to be $\sim$0.05\%/$\upmu$W in the linear operating regime of our superconductor.  We note that, particularly in the CW regime, the optical pump can have a dramatic effect on our microwave resonator linewidth. This is largely mitigated in the pulsed regime by maintaining a low average optical power. More details can be found in Extended Data Fig. \ref{fig:mw-nonlinearity}. In current experiments, the peak conversion efficiency measured both in the microwave $\rightarrow$ optical (MW$\rightarrow$ Opt.) and optical $\rightarrow$ microwave (Opt$.\rightarrow$ MW) conversion regimes is $\eta\approx 1.18$\% for  -13.8 dBm (44.2 $\upmu$W) of on-chip optical power (Fig. \ref{fig:perf}a). We note that this is a 400-fold improvement over our previous result \cite{holzgrafe.20} and a 500x improvement in conversion per $\upmu$W of pump power when compared to other integrated photonics EO transducers \cite{mckenna.20, Xu.21}. Our transducer also features  a bidirectional 3dB bandwidth  up to 30  MHz (Fig. \ref{fig:perf}b).  This is extracted by sweeping the frequency applied to the SSBM using a network analyzer. The roll-off in converted power measured at high optical pump powers corresponds to a distortion of our microwave resonator due to finite absorption of the superconductor resulting in nonlinear behaviour, as well as  drift in our triple-resonance condition during measurement.

From efficiency and linewidth measurements, we can extract the microwave-optical coupling rate ($g_{eo}$) and cooperativity (C) in order to understand the coherence of our system, given $C=\frac{4g_{eo}^2n_{-}}{\kappa_+\kappa_m}$. Here, $n_{-}=\frac{\kappa_{-,e}}{\kappa_-^2/2}\frac{P_{pump}}{\hbar\omega_-}$ is our intra-cavity photon number corresponding to the red optical pump for input power $P_{pump}$ and optical (extrinsic) decay rate $\kappa_{-(,e)}=70 $ (48.5) MHz. The loss rate of the blue optical signal field is described by $\kappa_{+}= 90$ MHz, and the pump-power dependent decay of the microwave signal field is described by $\kappa_m>7$ MHz (values reported in Extended Data Fig. \ref{fig:mw-nonlinearity}a). We find that the $g_{eo}/2\pi\approx945$ Hz and measure cooperativity up to 1.16\%. \\

\begin{figure}[h!]
    \centering
    \includegraphics{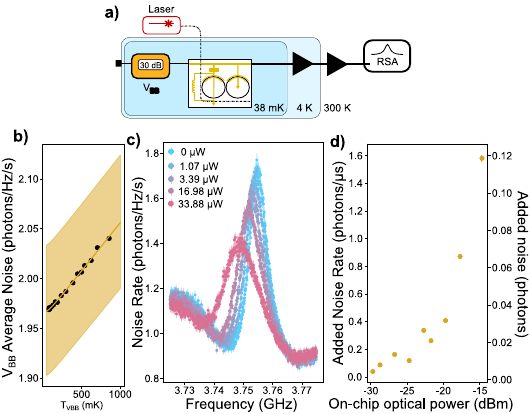}
    \caption{Evaluation of optically generated microwave noise in the transducer. (a) Measurement schematic. We send a continuous-wave (CW) optical pump to the transducer and measure the power spectral density ($P_{RSA}$) of the microwave signal generated on superconducting electrodes on a radio spectrum analyzer (RSA)  with a 1 kHz resolution bandwidth (RBW) averaging over 200 measurements. We note that the microwave signal is amplified using 4K and 300K amplifiers. We calibrate the link gain in the system using a variable temperature black-body source ($V_{BB}$) that is weakly thermalized to the mixing chamber plate\cite{hease.20}. (b) Calibration curve for the average output noise photon rate per hertz bandwidth of the $V_{BB}$ (black dots) that allows us to extract the link gain (G) and system added noise rate ($N_{sys}$) according to equation \ref{eqn:noise-calib}. The line of best fit is plotted in gold. The fit uncertainty is shaded. (c) Power spectral density (PSD) emitted by the device and converted into noise photon rate. Errorbars correspond to calibration uncertainty. The PSD measured with the laser beam blocked is plotted in blue.  (d) Added noise rate per microsecond (left axis) and noise photons emitted by the resonator during the resonator lifetime (right axis) for a given CW pump power. The errorbars correspond to the uncertainty due to calibration errors plotted in (b).  }
    \label{fig:noise}
\end{figure}

We next evaluate the optically-induced microwave noise generated by the transducer. In order to improve signal-to-noise ratio for these measurements, we characterize the device using the schematic shown in Fig. \ref{fig:noise}a and measure the output power spectral density (PSD) on a Real-Time Spectrum Analyzer (RSA).

We calibrate our link gain (G) and noise photons in our system ($N_{sys}$) using a variable temperature black-body ($V_{BB}$) source which follows the temperature distribution \cite{hease.20}
\begin{equation}
   P_{avg} \approx \frac{1}{2}G\hbar\omega_m BW coth\left(\frac{\hbar\omega_m}{2 k_B T_{V_{BB}}}\right)
    \label{eqn:noise-calib}
\end{equation} 
for source temperature $T_{VBB}$ and noise bandwidth BW. Since the black-body source is directly connected to the transducer, we assumed that the waveguide is also thermalized at temperature $T_{V_{BB}}$ (Fig. \ref{fig:noise}a). The black-body source is weakly thermalized to the mixing chamber plate, so the bath temperature remains less than 10\% of the temperature of the black-body source.  From Eqn. \ref{eqn:noise-calib}, we extract G=88.45$\pm$ 0.07 dB and $N_{sys}=112\pm$ 1.89 photons for a noise bandwidth BW = 30 MHz, selected to account for broadband added noise contributions. We attribute the high added system noise photon number to the high temperature of our BB source, added noise of amplifiers at 4 K and 300 K, and reflections from our 4 K amplifier due to the lack of isolation before the device.

After calibration, we characterize the transducer-added noise ($n_{add}$) under CW optical excitation by measuring the device output PSD (Fig. \ref{fig:noise}c). After each RSA measurement, we measure the transmission of our microwave resonator using a VNA before increasing the optical power in order to extract the microwave resonator lifetime at the measured pump power (Extended Data Fig. \ref{fig:mw-nonlinearity}b). We can then extract the added noise photon generation rate for a given CW optical pump power by integrating the measured PSD over our measurement bandwidth (30 MHz) and normalizing this to the PSD measured when the laser is shuttered (Fig. \ref{fig:noise}d, left axis). We also estimate added noise photons taking into account our microwave resonator bandwidth (Fig \ref{fig:noise}d, right axis), and observe low ($n_{add}<0.118\pm$0.002) transducer-generated noise photons for up to 33.88 $\upmu$W CW optical power. In general, we expect the Purcell-limited coherence time of a transmon qubit to be proportional to the fractional added noise. We note that, in general, we expect the fractional noise photons generated by the transducer to result in a corresponding penalty to the Purcell-limited coherence time of a qubit. 

\subsection{Transducer-qubit interconnect} 
\begin{figure*}[!htb]
    \centering
    \includegraphics{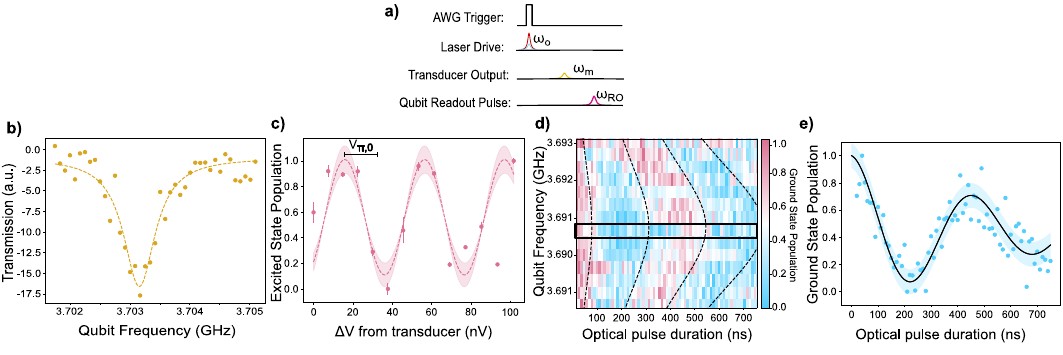}
    \caption{Optical control of superconducting qubits. (a) Pulse sequence for pulsed transducer-driven qubit measurements. A trigger is sent from the room-temperature qubit control electronics to trigger the optical pulses. Optical pulses at the transducer generate a microwave pulse of approximately equal pulse width. The pulse is sent to the qubit, after which a readout pulse is generated at room temperature, transmitted through the qubit readout resonator, and measured to extract the qubit state. See Sup. \ref{supp:setup} for full details. (b) Continuous-wave qubit spectroscopy using $\sim$328 nV microwave signal generated by the transducer. Flux biasing of the qubit is used to sweep the qubit frequency into resonance with the transducer signal. Population is read out via the dispersively-coupled readout resonator. (c) Optically driven qubit power Rabi oscillations using the pulse sequence described in (a) with 100 ns pulses. The optical on-chip pump power is kept constant at $P_{pk}=7\pm0.5$ $\upmu$W. The idler's optical power is swept from 400 to 500 nW, which corresponds to qubit drive voltages from 856 to 958 nV. Errorbars on population correspond to uncertainties when projecting datum onto a phase-quadrature axis. From our measurements, we fit the Rabi oscillation and evaluate a Rabi $\pi$-pulse amplitude $V_\pi\propto 20.28\pm 0.035$ nV satisfied for integer multiples of $V_{\pi,n}$. The fitting error is shade in pink. (d-e) Optically driven qubit time-Rabi oscillations. We fix the power in the optical pump and idler and sweep the width of the optical pulses to measure the corresponding change in qubit population. (d) We flux bias the qubit to sweep the detuning of the qubit from the optically-generated microwave tone in a characteristic chevron profile. Dotted lines are the theoretical prediction of Rabi frequency as a function of detuning following $\Omega_R'=\sqrt{\Omega_R^2+\delta^2}$ where $\Omega_R'$ is the Rabi frequency at detuning $\delta$ given resonant Rabi frequency $\Omega_R$. The population for different pulse widths near the optimal bias is shown in (e) where we measure a Rabi frequency of 2.27 MHz by fitting a Rabi oscillation with a time decay.  Errorbars on population correspond to uncertainties when projecting datum onto a phase-quadrature axis. The fitting error is shaded in blue.} 
    \label{fig:qubit}
\end{figure*}
We use the transducer to control a SC qubit using microwave signals generated by the transducer from optical inputs. Here, we drive an aluminum split-transmon qubit through the readout resonator. We operate the transducer in a classical red pumping regime to coherently convert an optical tone (idler) into a microwave tone in an interaction mediated by a strong optical pump using the pulse sequence depicted in Figure \ref{fig:qubit}a. Here, an arbitrary waveform generator (AWG) triggered by the qubit control board (Radio Frequency System on a Chip, RFSoC) is used to gate the optical pulses used to generate the microwave drive signal.  The RFSoC then sends a readout pulse to the qubit's readout resonator to measure the qubit state after excitation. 

We apply square optical pulses of length ($\tau_p$) much longer than the optical resonator linewidth ($1/\kappa_{\pm}\approx$ 15 ns, $\tau_p>70$ ns), which approximates an asymmetric Gaussian output microwave pulse (Sup. \ref{sup:shaping}). For each measurement, we use a 100 $\upmu$s repetition time between events to be consistent with baseline, traditional RF qubit measurements (see Extended Data Fig. \ref{fig:bringup}) while maintaining a low bath temperature (T$\approx$14 mK). 

We perform qubit spectroscopy and Rabi oscillation measurements. We first perform continuous wave (CW) qubit spectroscopy using a CW microwave tone generated by the transducer ($\omega_{MW} = 2\pi\times 3.703$ GHz) to excite the qubit population (Fig \ref{fig:qubit}b) in order to verify excitation transfer between the transducer and qubit and determine the transducer output power needed to excite the qubit.  

Next, we perform Rabi oscillation measurements in the power (Fig. \ref{fig:qubit}c) and time (Fig. \ref{fig:qubit}d-e) domains and extract parameters needed to impart a $\pi$-pulse to the qubit. For optically-driven qubit power-Rabi measurements, we keep a 100 ns optical pulse width ($\omega_{MW}=2\pi\times 3.71$ GHz) and uniform pump power ($P_{\omega_-}=7\pm5$ $\upmu$W) while linearly increasing the optical power in the blue sideband ($P_{\omega_+}$) arriving at the transducer to linearly increases the microwave power arriving to the qubit (Fig. \ref{fig:qubit}c). This results in a $\pi$-pulse voltage amplitude $V_\pi=$20.28 nV extracted by fitting the Rabi oscillations assuming a 50 $\Omega$ load at the qubit. Some output power instability (>3 dB measured before and after some successive measurements) due to power dependent drifts in our system motivated further measurements in the time-Rabi domain in order to maintain more uniform power at our single-sideband modulator and device. 

For optically-driven time-Rabi measurements (Fig. \ref{fig:qubit}d,e) we maintain uniform pump and idler power ($P_{pump} \approx 7\pm$ 0.5 $\upmu$W, $P_{idler} \approx 700\pm 150$ nW) and sweep the optical pulse width in order to change the width of the microwave pulse ($\omega_{MW}/2\pi = 3.6909$ GHz) arriving to the qubits. We attribute the lower operating frequency used here when compared to the frequency reported in Section 2 to thermal effects in our system or operation of the transducer away from the triple-resonance condition. 

The flux bias is swept to bias the qubit to different frequencies in order to measure the qubit excitation relative to the transducer frequency in a chevron-type measurement, where the Rabi frequency ($\Omega_r'$) changes as a function of detuning  $\delta$ by $\Omega_R'=\sqrt{\Omega_r^2+\delta^2}$ (dotted lines in Fig. \ref{fig:qubit}d). We fit the Rabi oscillations according to $P_{Rabi}=Ae^{-t/\tau}sin(2\pi t/(2T_\pi))+P_b$ to extract a $\pi$ shift to qubit for pulse width $T_\pi$, characteristic decay $\tau$, and offset $P_b$ and measure $T_\pi\propto$ 220 ns with $\Omega_R/2\pi=2.27$ MHz at the transducer frequency. Because the Rabi frequency is proportional to the incident microwave pump power, we can compare the measured Rabi frequency to the estimated transducer microwave output power (Methods M3.1) to estimate 46\% signal loss between the transducer and qubit. We attribute this high loss to microwave losses in the cables, circulators, filters, and directional coupler between the transducer and qubit due to a bandwidth mismatch between our operating frequency and cryogenic electronics.  

We measure a Rabi decay constant of $\tau=600\pm200$ ns. From transducer-generated added noise reported in Fig. \ref{fig:noise}d, we estimate the transducer operated at 7 $\upmu$W  peak optical power results in $n_{add}<0.02$ added noise photons during optically-driven qubit measurements, resulting in a 2\% transducer-noise penalty on the Purcell-limited qubit coherence time $\left(T_{MOQT}=\frac{1}{\gamma (n_{add}+1)}\right)$. We use RF measurements conducted with the RFSoC to estimate qubit decay constant $\gamma = 1.936$ kHz (Meth. \ref{meth:bi}). This allows us to estimate a transducer-limited coherence time of 82 $\upmu$s. This value is an underestimate, since the thermalized components in the interconnect will decrease the amount of thermal noise that reaches the qubits. The transducer-limited coherence time is much greater than our measured $\tau_{Rabi}=600\pm200$ ns and $T_2^*=710\pm5$ ns measured during baseline coherence measurements; we therefore conclude that our qubit coherence times are likely limited by other environmental noise sources.

\subsection{Discussions and Outlook}
We demonstrate a CEO-MOQT with high per-microwatt conversion efficiency in the linear regime of our SC resonator ($\%\eta/\upmu W=0.05$) driven with up to 44 $\upmu$W for a peak efficiency $\eta=1.18\%$, cooperativity $C=1.16\%$, and microwave-optical single-photon coupling rate $g_{eo}/2\pi=945$ Hz used for coherent optical-driving of a SC qubit.  Furthermore, we report low added microwave noise from operating the device, with $\bar{n}_{add}<0.12$ photons for up to 33 $\upmu$W on-chip optical power. More than an order of magnitude improvement in device performance could be made by reducing microwave losses, using a single-ended microwave resonator fabricated directly on the Si handle wafer (instead of on \ce{SiO2}), and by reducing the optical coupling losses to minimize scattering and leverage the full incident pump power on chip. Improvements to the microwave extraction efficiency and total link loss between the transducer and qubits will be essential to realize high-fidelity quantum information transfer in the future.

The increased optical bandwidth when compared to the microwave domain can be leveraged with a flux tunable microwave cavity \cite{xu.19}. Measurement improvements, such as a phase lock loop between the qubit readout electronics (or future transducers at the readout frequency) and the pump driving tone can allow for phase control of the qubits and more intricate gate operations between qubit nodes. By engineering the temporal profile of optical pump fields, we can generate shaped, symmetric, single microwave photons in order to achieve high efficiency photon transfer between the transducer and a qubit device (Sup. \ref{sup:shaping}). Finally, the low microwave noise generation when continuously pumping the device indicates that stronger optical pump fields can be used for conversion without significantly impacting qubit performance. Characterizing the link between the transducer and qubits with higher coherence time devices will allow us to better quantify the effect of the CEO-MOQT on SC qubit performance.

While the CEO-MOQT discussed in this article operates in the low cooperativity regime with modest conversion efficiency, the device performance shows potential for networking SC qubits using heralded remote-entanglement protocols when operating in the SPDC regime \cite{duan.01, krastanov.21}. This can enable the scaling of processors across multiple cryostats, eliminating volumetric constraints that currently exist in these systems. Combining this with the work presented here, as well as with recent demonstrations of optical readout of SC qubits \cite{arnold.23, van_thiel.23}, provides a path towards forming all-optical interfaces with SC qubits to enable the extension of SC quantum systems to enable large scale quantum processors \cite{lago-rivera.21,bhaskar.20}.

\section*{Acknowledgements}
The TWPA used in this work was provided by MIT Lincoln Laboratory. The qubit control code was built using the open source quantum instrument control kit (QICK). 

The fabrication of these chips was performed in part at the Center for Nanoscale Systems (CNS), a member of the National Nanotechnology Coordinated Infrastructure Network (NNCI), which is supported by the National Science Foundation under NSF Award no. 1541959. We thank Chiao-Hsuan Wang, Yat Wong, Mengzhen Zhang, Changchun Zhong, and Michael Haas for helpful discussions.

This work is supported by the AFRL under awards RCP06360 (H.K.W., J.H., D.B., N.S.), NSF under awards EEC-1914583 (H.K.W., N.S.), OMA-2137723 (D.B., C.J.X.) OMA-1936118, ERC-1941583, and OMA-2137642 (L.J.); DARPA under award HR01120C0137 (D.B., A.S.-A.); DoD under award FA8702-15-D-0001 (N.S.), DoE under award DE-SC0020376 (N.S.); AFSOR under awards FA9550-20-1 (D.Z.), FA9550-19-1-0399, FA9550-21-1-0209 (L.J.); ARO under awards W911NF-20-1-0248 (D.Z.), W911NF-23-1-0077, and W911NF-21-1-0325 (L.J.); and NTT Research, Packard Foundation under award 2020-71479 (L.J.)

H.K.W. acknowledges financial support from the National Science Foundation Graduate Research Fellowship under Grant No. 1745303. D.B. acknowledges financial support by the Intelligence Community Postdoctoral Fellowship. D.Z. acknowledges financial support from the HQI fellowship. E. B. acknowledges support from the National Science Foundation Graduate Research Fellowship under Grant No. 2141064. G.J. acknowledges financial support from the Natural Sciences and Engineering Research Council of Canada (NSERC).

\section*{Author Contributions Statement} 
J.H. and H.K.W. designed the transducer. D.B., H.K.W., J.H., and C.J.X. fabricated the transducer with help from E.B. and M.C. for Nb superconductor growth and A.S.-A. for fabrication development. Superconducting qubits were designed and fabricated by Rigetti Computing. H.K.W. built the experimental setup with help from J.H. and B.Y.  Measurements were designed by H.K.W., B.Y., and J.H. with help from N.S., D.Z., A.S., B.L., L.J., M.R., and M.L. The cryogenic measurement system was built and maintained by H.K.W., J.H., N.S., D.Z., and G.J.  Measurements and data analysis were completed by H.K.W. H.K.W. wrote the manuscript with contributions from all authors.

\section*{Competing Interests Statement}
J.H. and M.L. are currently involved in developing lithium niobate technologies at HyperLight Corporation. B.Y., S.P., E.S., B.L., and and M.R. are or have been involved in developing quantum computing technology at Rigetti Computing. The remaining authors declare no competing interests.


\bibliographystyle{sn-nature}

\section*{References}
\bibliography{optical-interconnect}

\renewcommand*{\thesection}{}
\renewcommand*{\thesubsection}{M.\arabic{subsection}}
\renewcommand*{\thesubsubsection}{M.\arabic{subsection}.\arabic{subsubsection}}

\setcounter{subsection}{0}%

\section*{Methods}
\subsection{Fabrication}\label{Meth:fab}
Devices are fabricated on 600-nm x-cut thin-film lithium niobate (TFLN) bonded on 4700~nm of thermally grown oxide on a silicon carrier (NanoLN). The waveguides are defined using negative-tone hydrogen silsesquioxane (HSQ) (FOx-16) by means of electron-beam lithography under 4x-multipass exposure (Elionix
F-125). The pattern is then transferred to LN through a physical etching process using  Ar\textsuperscript{+} ions to form a 325 nm ridge waveguide. We further confirm the thickness of the remaining film by means of an optical profiler (Filmetrics). We used a wet chemical process to clean the resist and redeposited material after the physical etching. The chip is then cladded with 1.4 $\upmu$m of inductively-coupled plasma enhanced chemical vapor deposition (IC-PECVD) SiO\textsubscript{2}. Windows are etched into the cladding oxide and waveguide slab before 40 nm of niobium is sputtered onto the chip. The microwave layer is then defined with positive tone propyleneglycol monomethyl etheracetate photoresist (SPR 700-1.0) by means of optical lithography (Heidelberg 150) and transferred to the niobium film using a \ce{CF4} physical etch. The etch depth is targeted to be 100nm to ensure that the microwave resonator is fully defined. 

\subsection{Transduction Efficiency}

\label{meth:eff}
We can characterize the on-chip conversion efficiency by comparing the output signal flux to the input idler flux. In the following section, we will describe measurement in the optical-to-microwave conversion regime: an equivalent process is used to characterize microwave-to-optical conversion, with MW ($\omega_m$)$\rightarrow$ opt. ($\omega_+$). 

For optical-to-microwave transduction experiments, we characterize the on-chip transduction efficiency by measuring the microwave signal power converted from our optical idler through the qubit transmission line. During these experiments, the qubits are detuned from the transducer frequency by >500 MHz.  We pump the device with a laser ($P_{pump}$) resonant with our red optical sideband ($\omega_-$) and optical idler at $\omega_+$ generated at room temperature ($P_{in, opt}$). After the device, we measure the microwave power generated at $\omega_{MW} $ at room temperature ($P_{out, MW}$). From this, we can find the input flux of signal $\Phi_{MW, opt.} = \frac{P}{\hbar\omega_{MW, opt}}$.  We measure optical losses at room temperature  with a powermeter (Thorlabs PM100D) calibrated against an integrating sphere (Thorlabs S140C). It is assumed that optical losses in our measurement network remained nominally constant throughout the experiment. We calibrate our microwave losses as described in Meth. \ref{meth:bi} in order to determine our input optical loss $\eta_{in, opt}$ and output optical loss $\eta_{out, opt}$. Finally, we can calculate our on-chip conversion efficiency ($\eta_{MOQT}$) by normalizing the idler and signal losses at room temperature to our input and output losses according to

\begin{align}
    \eta_{MOQT} & = \eta_{MW\rightarrow opt.} = \eta_{opt.\rightarrow MW} \label{eqn:bi1}\\
    \eta_{MW\rightarrow opt.} &= \frac{1}{\eta_{in, MW}\eta_{out, opt} }\frac{\Phi_{opt}}{\Phi_{MW}} \label{eqn:bi2}\\
    \eta_{opt.\rightarrow MW} &=\frac{1}{\eta_{in,opt}\eta_{out, MW} }\frac{\Phi_{MW}}{\Phi_{opt.}}
    \label{eqn:bi3}
\end{align}

The system losses are converted to decibel units (dB) and reported in Table I. The microwave input and output loss calibration, as well as the transducer-qubit link loss ($\eta_{MOQT-qubit}$) and estimated qubit RF control input loss ($\eta_{qubit-in}$) is described in Methods \ref{meth:bi}.
\begin{table}[]
    \centering
    \begin{tabular}{c|c}
 \textbf{Loss parameter}&\textbf{Link Loss (dB)}\\\hline
         $\eta_{in, MW}$&  -75.49\\
         $\eta_{out, MW}$& 
    +48.99\\
 $\eta_{in, opt.}$&-14.69\\
 $\eta_{out, opt}$&+5.51\\
 $\eta_{sys, MW}$&-26.5\\
 $\eta_{MOQT-qubit}$&-2.67\\
 $\eta_{qubit-in}$&-10\\
 \label{tab:loss}
 \end{tabular}
    \caption{Link input and output microwave and optical losses measured in our system as described in Methods \ref{meth:eff} and \ref{meth:bi}. We have amplifiers on the microwave and optical outputs, resulting in gain on the system outputs.}
\end{table}\\

\subsection{Microwave Loss Calibration} \label{meth:bi}
The input and output optical losses ($\eta_{in, opt.}, \eta_{out, opt.}$) can be measured at room temperature with fiber transmission measurements as described in the previous section. We cannot characterize our microwave losses with room temperature transmission measurements because the system is under vacuum and is sensitive to thermal fluctuations.  However, we can calibrate our microwave losses using low-power, bidirectional transduction measurements in order to determine the true on-chip conversion efficiency and our input and output microwave losses.

Additionally, we can measure out total microwave link loss ($\eta_{sys, MW}$) using microwave transmission measurements. This allows us to characterize our microwave input and output losses as $\eta_{sys, MW} (dB)=\eta_{in, MW} (dB)+\eta_{out, MW} (dB)$ and extract our link loss when we solve this with Eqn. \ref{eqn:bi1}-\ref{eqn:bi3} given that we only probe $\eta_{in, MW}$ during MW$\rightarrow$ opt. conversion and $\eta_{out, mw}$ for opt.$\rightarrow$MW. conversion.

\subsubsection{Transducer-qubit link loss} \label{meth:link loss}

The calibration described above allows us to extract the microwave losses before and after the transducer. It does not provide any information regarding the link loss between the transducer and qubit ($\eta_{MOQT-qubit}$). 

However, the transducer output microwave loss includes the loss through the qubit transmission line. Since the Rabi frequency ($\Omega_R$) is proportional to the total power ($P=\int P_{pk}dt$ for a peak power $P_{pk}$), we can use Rabi measurements in order to extract the link loss between the transducer and qubits. For a qubit with decay constant $\gamma$ at frequency $\omega_q$, the Rabi frequency is given by \begin{equation}
    \Omega_R = 2\sqrt{\frac{\gamma P}{\hbar \omega_q}}
\end{equation}

First, we must extract the qubit decay constant $\gamma$ using the RF calibration from the RFSoC. Here, Gaussian pulses are generated (ie, $P = \int P_{pk}dt\rightarrow\sqrt{2\pi\sigma^2} P_{pk}$) for pulse width $\sigma$ at room temperature before being sent to the fridge. For RFSoC time rabi measurements, we measure a room temperature pulse amplitude of 179.69 $\upmu$V and estimate overall -10 dB link loss between the RFSoC and qubits (including 3 dB propagation losses, 58 dB amplification, and 66 dB attenuation). This allows us to extract a qubit decay constant of $\gamma/2\pi =1.936$ kHz. 

We use the extracted decay constant to estimate the transducer-qubit link loss ($\eta_{MOQT-qubit}$) by comparing the measured Rabi frequency ($\Omega_{R, meas}$) to the expected Rabi frequency ($\Omega_{R, expected})$ for the normalized power we have measured being emitted at the transducer $(P_{MOQT, MW}$) since
\begin{equation}
    \frac{\Omega_{R, meas}}{\Omega_{R,expected}}=\sqrt{\frac{P_{qubit}}{P_{on-chip,MW}}}
\end{equation}
for microwave power at the qubit $P_{qubit}$, where $P_{qubit} = \eta_{MOQT-qubit}P_{on-chip, MW}$ or  
\begin{equation}
    \eta_{MOQT-qubit} = \left( \frac{\Omega_{R, meas}}{\Omega_{R,expected}}\right)^2
\end{equation}
From our microwave loss calibration, we expect -114.5 dBm emitted by the transducer in an asymmteric Gaussian of width $1/\tau$ for square optical pulses of length $\tau$ and a tail determined by the microwave decay $\kappa_{MW}$. Here, we approximate the total power $\Sigma P = \int P dt = P(\sqrt{\pi}\tau + (1-1/\kappa_{MW}))$ . This results in an expected Rabi frequency of $\Omega_{R, meas}=2\pi\times4.20$ MHz, which we compare to our measured Rabi frequency $\Omega_{R, meas}=2\pi\times2.27$ MHz. We estimate 2.67 dB (46\%) loss between the transducer and qubits. We attribute this largely to propagation losses through system components (coaxial cable, filters, circulators, and a directional coupler) between the transducer and qubits. 
For off-resonant drive, like the chevron measurement in Fig. \ref{fig:qubit}d, we expect to measure a rolloff in Rabi frequency 
\begin{equation}
    \Omega_{R}^{'} = \sqrt{\Omega_R^2+\delta^2}
\end{equation}
where $\delta$ is the detuning between the qubit frequency and driving tone. 

\subsection{Transducer Measurement}\label{meth:transducer}
For independent transducer measurements, a Santec TSL-550 CL band laser operated around 1530 nm is sent to a single-sideband electro-optic modulator.  For optical-to-microwave conversion experiments, the modulator is weakly driven by a vector network analyzer (VNA) at the transducer microwave resonator frequency to generate an optical sideband.  For microwave-optical conversion experiments, the VNA output is switched to the dilution refrigerator input and used to apply a weak microwave idler to the transducer.  The optical pump(s) can be switched to an acousto-optic modulator (AOM) and electro-optic modulator (EOM) for phase locking the optical pump laser to the optical resonator using the Pound-Drever-Hall technique. A DC bias is applied to the transducer bias capacitor by applying a DC voltage to a cryogenic bias-tee (QMC-CRYOTEE-0.218SMA) using a Keithly 2400 power supply.

\subsection{Noise Measurement}\label{meth:noise}
The noise measurement was performed after the qubits were removed from the setup and bypasses cryogenic circulators, amplifiers, and filters below 4K, with amplification stages at 4K and room temperature. The variable black-body source ($V_{BB}$) is constructed from a 30 dB attenuator thermalized to a copper block. The copper block houses a cartridge heater (Lakeshore HTR-50) and temperature sensor (Lakeshore RX-102B-RS) that allows us to apply voltage-dependent temperature changes to our microwave input line. The $V_{BB}$ is weakly thermalized to the mixing chamber plate using a polymer spacer, which allows us to heat the waveguide to $\sim$1K while maintaining a mixing-plate temperature $<100$ mK. The high added base temperature results in large added noise photon number. The heater is terminated for optically pumped noise measurements. Off-resonant laser power is sent through the device via a fiber array coupled to on-chip grating couplers.

\subsection{Qubit Control } \label{meth:qubit}
Qubit control was implemented using the Xilinx ZCU111+ Radio Frequency System on Chip (RFSoC) with control built using the open-source Quantum Instrument Control Kit (QICK). Qubits could be characterized independently from the transducer by sending in control pulses on a directional coupler after the transducer chip to give a baseline for qubit measurements using 50 ns pulses with 100 $\upmu$s repetition rate averaged over 1000 experimental shots. This allowed us to establish a baseline for qubit lifetimes ($T_1$), coherence times ($T_2^*$), and RF readout fidelities. 

\subsection{Optically-Driven Qubit Measurements} \label{meth:inter}
In transducer-driven power-Rabi measurements, output power was swept from -108.33 to -107.36 dBm for 100 ns pulse width measurements and converted to voltage units assuming a 50 $\Omega$ load. Here, measurements were averaged over 1000 experimental shots.

For transducer driven time-Rabi measurements, output power was held fixed at $-115.8\pm0.4$ dBm.  Power was regularly checked on a VNA between each frequency flux bias in order to verify that the transducer bias had not drifted. Here, measurements were averaged over 5000 experimental shots.

\textbf{Data Availability}
Data shown in the main text and supplementary material is available upon request.

\textbf{Corresponding Authors} 
Correspondence and requests for materials should be addressed to Hana K. Warner\\ (hwarner@g.harvard.edu) or Marko Lon\v{c}ar (loncar@seas.harvard.edu).

\textbf{Supplemental Information} 
Supplemental information is available. 

\newpage

\renewcommand*{\thesection}{}
\renewcommand*{\thesubsection}{S.\arabic{subsection}}
\renewcommand*{\thesubsubsection}{.\arabic{subsubsection}}

\setcounter{subsection}{0}%

\section*{Supplementary Information}
\subsection{System Dynamics}
\label{sup:dynamics}
In this article, we describe an electro-optic transducer used to drive a superconducting qubit. In the transducer, the microwave and optical domains are coupled by the Pockel’s electro-optic effect to mediate a three-wave mixing interaction 
\begin{equation}
    \frac{H_0}{\hbar}=\omega_{a_1}\hat{a_1}^\dagger \hat{a}_1+\omega_{a_2} \hat{a}_2^\dagger \hat{a}_2+\omega_b \hat{b}^\dagger\hat{b}
    \end{equation}
    \begin{equation}
        \frac{H_I}{\hbar} = g_{eo}(\hat{a}_1 \hat{a}_2^\dagger \hat{b}+\hat{a}_1^\dagger \hat{a}_2\hat{b}^\dagger)
\end{equation}
Where $g_{eo}$ describes the single-photon coupling rate between the microwave ($\hat{b}$) and optical modes, $\hat{a}_{\{1,2\}}$. 

In our device, two hybridized optical resonators are capacitively coupled to a microwave LC resonator. The resulting eigenmodes $a_\pm$ are delocalized from the waveguide modes.  An optical pump field on the red optical mode mediates the conversion between the blue optical mode and microwave mode according to the reduced Hamiltonian 
\begin{equation}
        \frac{H_{red}}{\hbar}=-\Delta_+\hat{a}^\dagger_+\hat{a}_++\omega_m\hat{b}^\dagger \hat{b}+g(\hat{b}^\dagger \hat{a}_+ +\hat{ b}\hat{a}^\dagger_+)
\end{equation}
in a frame rotating at laser frequency $\omega_L\approx\omega_-$, where $\hat{a}_\pm$ are the eigenmodes of the hybridized optical resonator at frequencies $\omega_\pm$, $\Delta_+ = \omega_L-\omega_+$ is the detuning between the laser field $\omega_L$ and the blue optical mode $\omega_+$, $\hat{b}$ is the microwave mode at frequency $\omega_m$, and $g$ is the pump enhanced coupling term 
\begin{equation}
    g=\sqrt{\bar{n}_-}g_{eo}
\end{equation}
Where $\bar{n}_-$ describes the number of red pump photons defined by 
\begin{equation}
    \bar{n}_- = \frac{\kappa_{e,-}}{\Delta_-^2+(\kappa_-/2)^2}\frac{P_{pump}}{\hbar\omega_L}
\end{equation}
For total (extrinsic, intrinsic) loss rate $\kappa_{-(e,i)}$, detuning $\Delta_- = \omega_L-\omega_-$, and pump power $P_{pump}$.

The microwave signal generated by our transducer at $\omega_m$ is sent to a superconducting qubit ($\hat{q}$) via coaxial cable to allow for modular integration between the two sub-systems. 

We use a flux-tunable split-transmon qubit dispersively coupled to a readout resonator ($\omega_{ro}$).  For a qubit-cavity coupling rate $g_{q-ro}$ and qubit (cavity) loss rates $\kappa_{q (ro)}$, the system can be described by  
\begin{equation}
    \frac{H_{q-ro}}{\hbar} = \left(\omega_r + \frac{g^2}{\Delta}\hat{\sigma}_z\right)\left(a^\dagger a + \frac{1}{2}\right) + \frac{\omega_a}{2}\hat{\sigma_z}
\end{equation}

in the dispersive coupling regime where $g_{q-ro}>>\kappa_q, \kappa_{ro}$.  Here, $\Delta_q = \omega_{ro}-\omega_q$ is the detuning of the readout resonator from the qubit frequency and $\sigma_z$ is the Pauli z operator describing population exchange between the qubit ground and excited states. This allows us to measure the excitation generated by the resonant transducer signal by measuring the the qubit state-dependent frequency shift in the readout resonator.
\subsection{Electro-optic Coupling Rate}
\label{sup:eo}
We can derive the single photon coupling rate in the transducer from our three-wave mixing interaction given
\begin{equation}
    H_I=g_{eo}(\hat{a}_1\hat{a}_2^\dagger \hat{b}+\hat{a}_1^\dagger \hat{a}_2 \hat{b}^\dagger)
\end{equation}
so the interaction of an optical photon $\hat{a}_1$ with microwave photon $\hat{b}$ produces optical photon $\hat{a}_2$ such that $\omega_{a_2}=\omega_{a_1}+\omega_b$ and the decay of photon $\hat{a}_2$ to $\hat{a}_1$ and $\hat{b}$ with the same energy conservation requirement. 

Here, $g_{eo}$ is derived from the interaction energy due to microwave field modulating the index of refraction of the optical resonator due to electro-optic effect, which changes $\sim 1/\epsilon$ for the applied microwave field.  Thus, we describe the interaction in terms of the inverse permittivity matrix $\eta$ and optical displacement field $D_0$. For our system, this is defined by the crystal axis of the LN waveguide and resulting electric permittivity and the electric field strength in the waveguide described by

\begin{align}
    H_I&=\frac{1}{2\epsilon_0}\int_{LN}dV\sum_{ij}\Delta\eta_{ij}D_{oi}D_{oj}\\
    &=\frac{\epsilon_0}{2}\int_{LN}dV\sum_{ij}\Delta\eta_{ij}\epsilon_{ii}\epsilon{jj}E_{oi}E_{oj}
\end{align}
(which follows from $D=\epsilon E+P$)

Assuming fields are aligned along the crystal coordinates and substituting the electro-optic coefficient tensor $r$ for $\Delta\eta_{ij}=\sum_{ijk}r_{ijk}E_{\mu k}$, and defining the microwave and optical fields as $E_\mu=(e_bb+h.c)$, $E_0=(\hat{e}_{\hat{a}_1}\hat{a_1}+e_{\hat{a}_2}\hat{a}_2+h.c)$ it follows that \begin{equation}
    H_I=\epsilon_0\int_{LN}dV\sum_{ij}\epsilon_{ii}\epsilon_{jj}r_{ijk}(\hat{e}_{a_1i}+\hat{e}^*_{\hat{a}_2j}\hat{e}_{bk}\hat{a}_1\hat{a}_2^\dagger 
    \hat{b}+h.c)N_{\hat{a}_1}N_{\hat{a}_2}N_{\hat{b}}
\end{equation}
As derived in \cite{mckenna.20}. Here, only energy conserving terms are maintained.  Each electric field is normalized to zero point energies such that \begin{equation}
    N_m=\sqrt{\frac{\hbar\omega_m}{2\epsilon_0\int dV\sum_{ij}\hat{e}_{ij}\hat{e}_{mi}\hat{e}_{ij}}}
\end{equation}
In our device, the primary field contribution is along the $r_{33}$ TFLN crystal axis, so we can simplify the electro-optic tensor such that
\begin{equation}
    g_{eo}\approx\epsilon_0 n_e^4 r_{33} \int_{LN} dV \hat{e}_{\hat{a}_1} \hat{e}^*_{\hat{a}_2} \hat{e}_{\hat{b}} N_a N_b N_c
\end{equation}

allowing us to simulate the expected microwave-optical coupling rate $g_{eo}$ by examining the microwave field in the waveguide in a COMSOL multiphysics simulation.  

In this work, we describe a ring resonator with circumference $l$ (resonant condition $l=\frac{k\lambda_0}{n}$, where $k$ is the mode index, $\lambda_0$ is the free space wavelength, and $n$ is the (effective) refractive index).  

resulting in optical modes with frequencies $f_{0,k}=\frac{kc}{ln}$.  Changes to index of refraction $n\rightarrow n+\Delta n$ results in \begin{equation}
    f'_{0,k}=\frac{kc}{l(n-\Delta n)}\approx f_{(0,k)}\left(1-\frac{\Delta n}{n}\right)
\end{equation}
An electric field $E$ inside the optical waveguide linearly shifts the refractive index due to the Pockel's electro-optic effect, so for electric field applied along one direction \begin{equation}
    n_{LN}(E)\equiv n_{LN}-\frac{1}{2}rn_{LN}^3E
\end{equation}
where $r=-\frac{2}{n^3}\frac{dn}{dE}|_{E=0}$ is the electro-optic coefficient.  Along the z-axis of LN, $r=r_{33}\approx30$ pm/V, $n_{LN}=n_e=2.14$, and the electric field is highly uniform within the waveguide so we neglect the cross sectional variation of $E$ yielding the electro-optic susceptibility\begin{equation}
    \frac{G_i}{2\pi}=\frac{n_e^2r_{33}f_0\alpha\Gamma}{2|V/E|}
\end{equation}
where $\alpha$ is the fraction of the ring covered by electrodes, $\Gamma=\frac{\delta n}{\delta n_{LN}}\frac{n_LN}{n}$ is susceptibility of effective mode index in waveguide to changes in refractive index (this is essentially a confinement factor, with $\Gamma\approx1$)

This results in a vacuum coupling strength\begin{equation}
    \frac{g_i}{2\pi}=G_iV_{ZPF}
\end{equation}
From here, we can define the zero-point fluctuations $V_{ZPF}=\langle0|V^2|0\rangle$ of the LC resonator by finding the RMS voltage in an LC resonator with capacitance C and energy equal to vacuum energy of the resonator, \begin{equation}
    CV^2_{ZPF}-\frac{1}{2}\hbar\omega_m
\end{equation}
yielding \begin{equation}
    g_i=\frac{n_e^2r_{33}w_0\alpha\Gamma}{2|V/E|}\sqrt{\frac{\hbar\omega_m}{2C}}
\end{equation}

We compare a fully cladded device (Extended Data Fig. \ref{fig:g0}a) to a device where we etch down the cladding and slab to create a "plateau" around the waveguide and gate the electrodes directly with the $z$ axis of the \ce{LiNbO3} crystal (Extended Data Fig. \ref{fig:g0}b). We simulate the electric field response in COMSOL multiphysics and infer device parameters to predict a 40\% improvement in $g_{eo}$ for a "plateau etch" geometry when compared to a fully cladded device. This is due to the stronger electro-optic field overlap in the waveguide and avoiding dielectric losses in the \ce{SiO2} cladding (Fig \ref{fig:g0}c).

\subsection{Microwave Pulse Shaping}
\label{sup:shaping}
When we operate in the pulsed transduction regime, we can "shape" our output pulses by controlling our input pump ($\omega_-$) and idler ($\omega_{+,MW}$) pulses. We examine this in the optical-to-microwave conversion regime, where pulse shaping is useful for efficient capture of the microwave signal by the qubit.  In this case, our microwave-optical coupling rate $G_{eo} = g_{eo}\sqrt{\bar{n}_-}$ develops a time dependence since the circulating number of pump photons $\bar{n}$, has a time dependence set by the pump envelop ($\bar{n}\sim P_{pump}(t)$).  Thus, we can re-solve the system equations described in Sup. \ref{sup:dynamics}, but now considering that $P_{pump}\rightarrow P_{pump}(t)$ and that the system is no longer in steady state, taking the approach described in \cite{cirac.97}.

Assuming we're still in a resonant regime where the blue-pumping Hamiltonian (spontaneous parametric down conversion) is suppressed, our Hamiltonian (rotating at the laser frequency) becomes \begin{equation}
    H = \Delta_+ \hat{a}_+^\dagger\hat{a}_+ + \omega_m \hat{b}^\dagger\hat{b} + g(t)(\hat{b}^\dagger\hat{a}_++\hat{b}\hat{a}_+^\dagger)
\end{equation}
where the system no longer assumes any modes are in steady state ($\hat{a}_+\rightarrow \hat{a}_+(t)$ and $\hat{b}\rightarrow\hat{b}(t)$). 

We consider the case described in this work, where input idler and pump are square pulses ($a_{+,in}(t) = A_+, 0<t<\tau$, $a_{-,in}(t) = A_-, 0<t<\tau$), respectively, yielding a square time dependence for the electro-optic coupling rate $g(t)=g_{eo}\sqrt{\bar{n}(t)} $. The two optical pulses are captured by our optical resonators according to $\hat{a}_{in,\pm} = 1-e^{-\kappa_{\pm,e}t}$ where they circulate for time $1/\kappa_\pm$. During this time, the microwave photon is generated and emitted from the cavity according to $e^{-\kappa_m(t-\tau)}$, where we assume simultaneous loading and decay of the square pulse. For width $\tau$ in the case $\tau>>1/\kappa_{\pm}$ and $\kappa_{\pm}>\kappa_m$, we expect the system to approximate an asymmetric Gaussian with a lopsided tail set by $\kappa_m$ (Extended Data Extended Data Fig. \ref{fig:pulse_shapes}).  In future work, different pump pulses, such as a sawtooth wave or pulses resembling the time-reversed pulse capturing profile, can be applied to improve the symmetry of our emitted microwave signal and signal capture efficiency. 

\subsection{Measurement Setup}
\label{supp:setup}
The measurement setup is depicted in Extended Data Fig. \ref{fig:full_schem}, with simplified schematics for Opt.$\rightarrow$MW transduction depicted in Extended Data Fig. \ref{fig:full_schem}a and MW$\rightarrow$ Opt. transduction depicted in Extended Data Fig. \ref{fig:full_schem}b.  For MW$\rightarrow$Opt. transduction (Fig \ref{fig:full_schem}a), a strong pump laser (red) at $\omega_-$ is modulated by a vector network analyzer (VNA) at a single sideband modulator (SSBM) near the microwave resonator frequency (yellow) to generate a weak optical sideband at $\omega_+$ (blue). The field field can then be modulated by an acousto-optic modulator (AOM) driven by an arbitrary waveform generator (AWG) to pulse the input optical fields.  These are then amplified by an erbium doped fiber amplifier (EDFA) and filtered by a band-pass filter (BPF) to compensate for losses in the bulk modulators. The optical fields are then coupled into the devices with a fiber grating array and interact at the transducer to generate a microwave tone that can be measured on the VNA after amplification at 15 mK by a Josephson traveling wave parametric amplifier (TWPA), 4K by a high electron mobility transistor (HEMT), and room temperature by low noise amplifiers. In (b), a strong pump (red, $\omega_-$) is sent to the AOM in order to pulse, amplify, and filter the field before interacting with a microwave tone (yellow) generated by the VNA at the transducer. The beatnote between up-converted optical field (blue) and pump are measured at a calibrated detector to measure the converted optical power.

The full experiment setup is depicted in Extended Data Fig. \ref{fig:full_schem}c. For independent transducer measurements, qubit components are treated as loss.  A directional coupler before the qubits allows us to characterize the qubits independent of the transducer. 


For independent qubit measurements, a qubit driving pulse at the qubit frequency is generated by the RFSoC and transmitted to the qubit via a directional coupler (QMC-CRYOCOUPLER-20) after filtering and amplification. The qubit can be DC flux biased between 3.25 and 4.5 GHz using a Keithly 2400 power supply. A readout pulse can then be generated at the qubit readout resonator frequency to readout out the qubit state after amplification.  We amplify first at base temperature using a TWPA, then at 4K using a HEMT, before finally amplifying at room temperature in two steps using low noise amplifiers before (ZX60-83LN-S+) and after (RF Bay LNA-2500) mixing our signal down to our board frequency using a double balanced mixer (ZX05-63LH-S+). 

For transducer-driven qubit measurements, the readout pulse is generated at the RFSoC and directed to the readout resonator on a separate transmission line and collected at the RFSoC after amplification and analog processing.  The transduced tone and the readout pulse are separated at the output of the dilution refrigerator to measure optical-microwave transduction, using a vector network analyzer (VNA), and qubit excitation, using the RFSoC, simultaneously.  Each measurement is averaged over many (1000-10,000) experiment shots.  The outputs are split on a 3dB coupler (QMC-CRYOCOUPLER-20) to allow for simultaneous monitoring of the transducer and qubit outputs. 

We use 40 dB of isolation (QMC-CIRC-4-12) between our system components (transducer, qubits, TWPA). An inner/outer DC block (Narda-Miteq 4563) prevents DC flux leakage from the transducer and qubit bias. 

We perform qubit power-Rabi measurements by keeping a fixed optical pulse width and pump power ($\omega_-$) while linearly increasing the optical power in the blue sideband ($\omega_+$) arriving at the transducer. A long relax-delay compared to the optical pulse width allowed us to maintain a low bath temperature (T$\approx$14 mK). We report measurements for a 100 ns optical pulse, which we expect to result in a corresponding microwave pulse width.  

For transducer-driven time-Rabi measurements we keep fixed power in the optical pump and idler  and sweep the optical pulse width in order to change the width of the microwave pulse arriving to the qubits. The transducer output power is fixed to -115.8 $\pm$ 0.4 dBm. The flux bias is swept to bias the qubit to different frequencies in order to measure the qubit excitation relative to the transducer frequency in a chevron-type measurement. This allows us to ensure that the qubit is biased at the transducer frequency.

\subsection{Transducer Packaging}
\label{sup:packaging} The transducer is enclosed in a copper box and wirebonded to a 50 $\Omega$ waveguide. We couple into the device with an optical grating coupler array using three axis attocube controllers on the mixing chamber plate of our dilution refrigerator.

Optical grating couplers (Extended Data Fig. \ref{fig:gcs}a) are created on chip by etching a 15$\times$15 $\upmu$m pad with teeth with period $\Lambda=750$ nm with separation of $\delta=562$ nm. This results in a phase matching condition of $k_2=k_1sin\theta + \frac{2\pi n_{eff}}{\Lambda}$ where $k=2\pi/\lambda$ is the wave-vector in the material propagating at angle $\theta=8^\circ$. A cross section of the grating couplers is drawn in Extended Data Fig. \ref{fig:gcs}b. In our system, transmission is optimized for operation at 1560 nm with a 30 nm 3 dB bandwidth (Fig \ref{fig:gcs}c).

We etch a niobium alignment pattern to facilitate blind coupling in our dilution refrigerator (Extended Data Fig. \ref{fig:gcs}a). The niobium ground plane is etched away between the grating couplers and device region, leaving $\sim$ 7.5 mm of exposed silicon dioxide between the optical coupling and transducer regions.  We expect this, combined with enclosed packaging and the low kinetic inductance of niobium, can help mitigate optically-induced noise generation in our superconductor. Optical insertion loss measured at peak grating coupler transmission was $\sim$ 4.5 dB/facet; however, the mode pair described here was slightly outside of the grating coupler bandwidth, resulting in an insertion loss of $\sim 10$ dB/facet.

The qubit devices are wirebonded to a PCB and packaged in a puck that sits inside of three levels of magnetic shielding thermalized to the mixing chamber plate.


\subsection{RF Qubit Characterization}
\label{sup:qubit}
The qubit is initially characterized in the RF domain using the Xilinx ZCU111 RFSoC.  Measurement control was written using the Quantum Instrument Control Kit (QICK). With the exception of time-Rabi measurements (where pulse width was swept), we use 50 ns Gaussian pulses with 100 $\upmu$s repetition rate averaged over 1000 shots. Measurement fidelity was evaluated as F=71.2\% using successive single shot $\pi$- and 0-pulse measurements.

For readout resonator spectroscopy, we measure the transmission of a readout pulse through the qubit measurement lines to find the resonance frequency (Extended Data Fig. \ref{fig:bringup}a).  We measure at swept output readout pulse powers to ensure that we are operating in the dispersive coupling regime. Similarly, we perform qubit spectroscopy by transmitting successive driving pulses at the qubit frequency and readout resonator frequency to readout the qubit state (Extended Data Fig. \ref{fig:bringup}b). We then complete qubit power Rabi measurements to find the qubit $\pi$-pulse voltage, and then perform lifetime ($T_1$) and coherence time $(T_2^*)$ measurements by sending $\pi$-pulses and successive $\frac{\pi}{2}$-pulses at the qubit frequency.  We extract a $T_1\approx 8$ $\upmu$s and $T_2^*$ $\approx 800$ ns.  

Finally, we can measure qubit RF-driven time-Rabi oscillations and compare them to the case of optically-driven time-Rabi oscillations.  Here, for similar on chip microwave powers, we see a larger characteristic time constant decay $\tau_{Rabi}\approx$ 220 ns in RF-driven time-Rabi oscillaiton measurements, compared to a characterstic decay of $\tau_{Rabi}\approx 800$ ns for the optically-driven time-Rabi oscillations.  This suggests in the optically-driven time-Rabi case, our decay time is limited by our qubit $T_2^*=800$ ns, whereas for the RF-driven case there is added noise in our room temperature measurement chain that could be reducing the coherence time of our qubits for long RF pulses.

\subsection{Optically-induced Decoherence}
\label{sup:decoh}
We characterize the qubit performance while optically pumping the transducer with a single strong pump field (not generating coherent microwave signals).  This allows us to characterize qubit performance ($T_1$, $T_2^*$, RO fidelity) under the effect of thermal bath heating or microwave noise generation due to our optical pump. 
 
We characterized the qubits while flux biased to the transducer frequency ($f_{ceo}=$3.71 GHz) and at the maximum frequency ($f_{max}=$4.57 GHz) while the laser was parked on resonance with the red mode of the transducer and while detuned from resonance.  We observed little effect of the transducer pump on the qubit lifetimes under optical pumping, and qubit lifetimes (Extended Data Fig. \ref{fig:decoh}a) and coherence time (Extended Data Fig. \ref{fig:decoh}b)  remained fairly consistent with the qubit performance independent of the transducer. Our measurements therefore indicate that the optical powers required to drive the transducer do not limit the $T_1$ of our qubits (at least to $\sim$9$\upmu$s). It is noted that this baseline characterization occurred at 0V flux bias, which corresponds to a qubit frequency of 4.2 GHz.

Baseline qubit readout fidelities before connecting the transducer and qubit with coaxial cables were measured as $F\approx 72$\% with successive single shot measurements. The transducer and qubits were then connected in a later cooldown, where we characterized qubit performance while optically pumping the transducer. We measured a similar base readout fidelity ($F\approx 70\%$) when the two systems are combined. We see a decay of the RO fidelity above -25dBm on chip optical pumping power (Extended Data Fig. \ref{fig:decoh}c). As a result, we targeted average optical powers $P_{avg}\approx-30$dBm for our combined experiment in order to keep readout fidelity as high as possible.  

It is noted that readout fidelity in this system was highly dependent on the TWPA pump frequency, mixer image frequency, and amplifier bias, indicating system degradation due to room temperature qubit and TWPA control electronics. Further improvements to the system are needed to verify whether the transducer could impose further challenges to qubit readout.

\renewcommand{\figurename}{Extended Data Fig.}
\setcounter{figure}{0}%

\twocolumngrid
\section*{Extended Data}

\begin{figure}[!ht]
    \centering
    \includegraphics{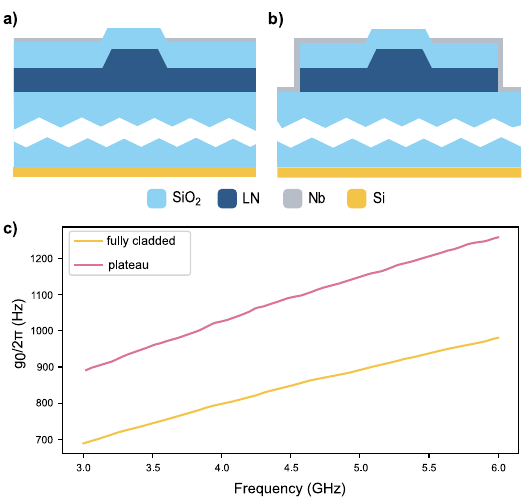}
    \caption{Electro-optic coupling rate. We simulate the microwave-optical single photon coupling rate, $g_0$, for (a) fully cladded and (b) plateau etch electrode geometries.  (c) We see a ~1.4x improvement in single photon coupling rate when comparing the plateau etch to a fully cladded device. }
    \label{fig:g0}
\end{figure}

\begin{figure}[!ht]
    \centering
    \includegraphics{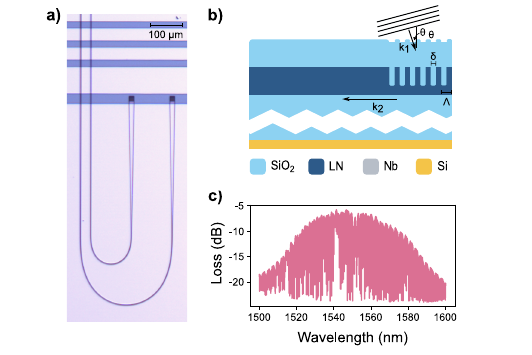}
    \caption{Transducer grating couplers. (a) Optical micrograph of grating couplers etched in TFLN cladded with \ce{SiO2}. A Nb alignment grating is created around the couplers in order to help with coupling inside of the dilution refrigerator. (b) Cross section of grating couplers for transducer material stack. Light is coupled into the waveguide according to the phase matching condition $k_2=k_1sin\theta + \frac{2\pi n_{eff}}{\Lambda}$ where $k=2\pi/\lambda$ is the wave-vector in the material propagating at angle $\theta$ and gratings are described by pitch $\Lambda$ and separation $\delta$. (c) Loss per grating measured through our device. }
    \label{fig:gcs}
\end{figure}

\begin{figure*}[!ht]
    \centering
    \includegraphics{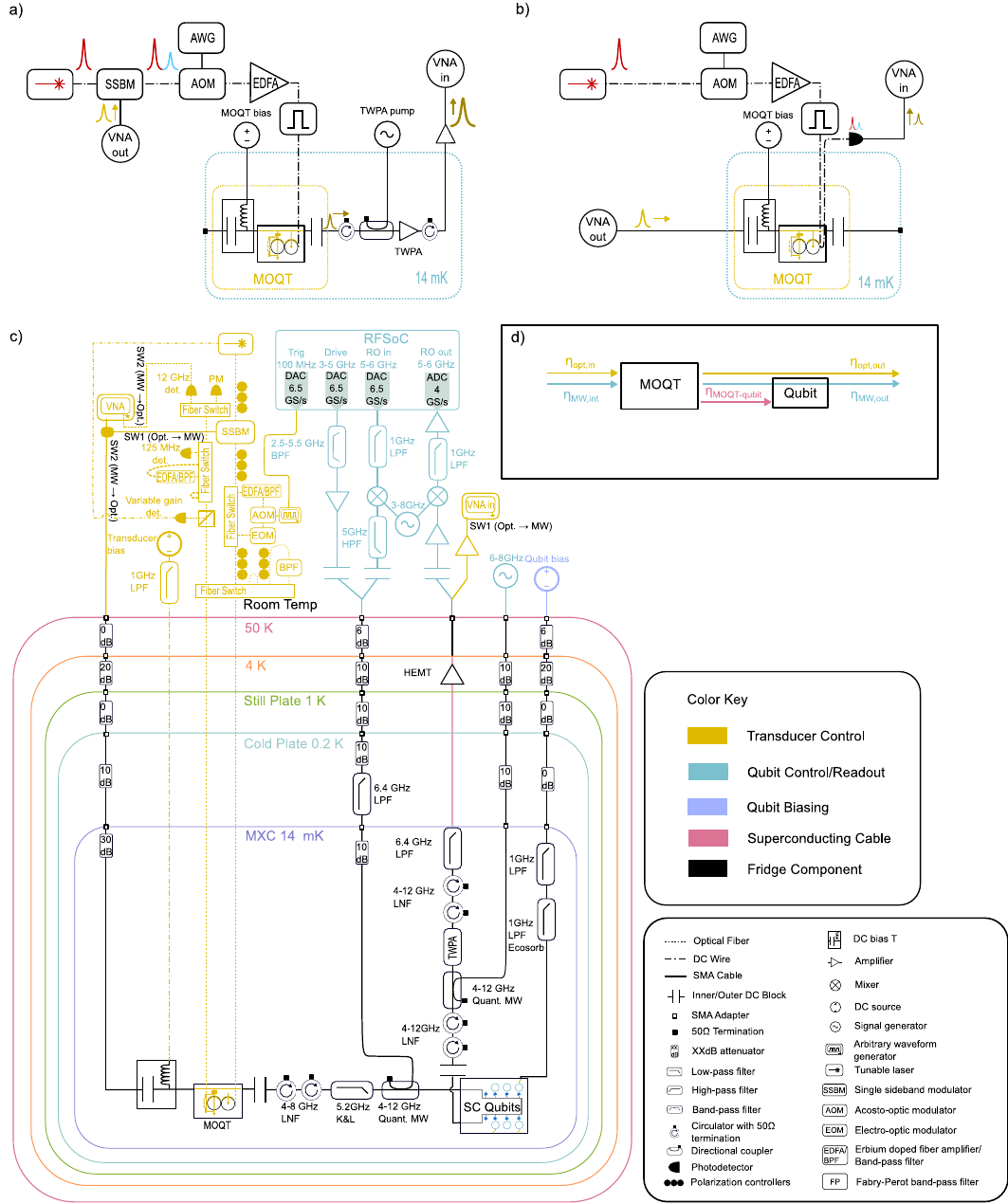}
    \caption{Schematic for CEO-MOQT and SC qubit measurements. (a) Simplified schematic for microwave$\rightarrow$optical conversion and (b) optical$\rightarrow$microwave conversion. Here, solid lines correspond to electrical cabling and dotted lines correspond to optical fiber.   (c) detailed schematic for the whole measurement system. Transducer components are highlighted in gold. Qubit components are highlighted in pink. (d) Definition of link losses ($\eta$) in the system.}
    \label{fig:full_schem}
\end{figure*}

\begin{figure}[!ht]
    \centering
    \includegraphics{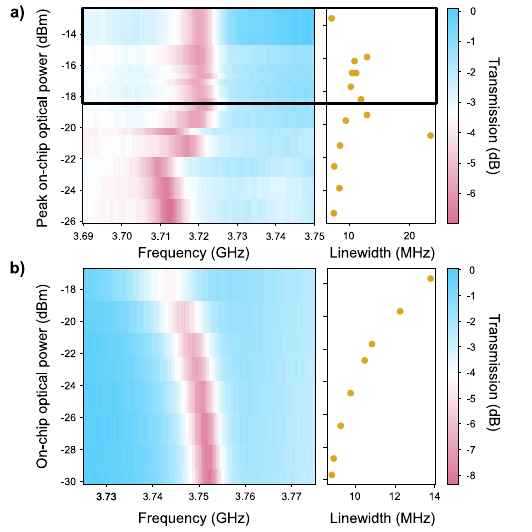}
    \caption{Microwave resonator spectra and linewidths (plotted in gold) for measured on-chip optical pump powers. (a) Microwave characterization during transduction efficiency measurements. Powers measured with pulsed optical pumps are enclosed in a black box. We selected our duty cycle for each pulsed pump power to maintain low thermal bath temperature, which allowed us to maintain a microwave linewidth $\sim$ 10 MHz for our pulsed transduction efficiency measurements. (b) Microwave transmission during noise measurements. }
    \label{fig:mw-nonlinearity}
\end{figure}

\begin{figure*}[!ht]
    \centering
    \includegraphics{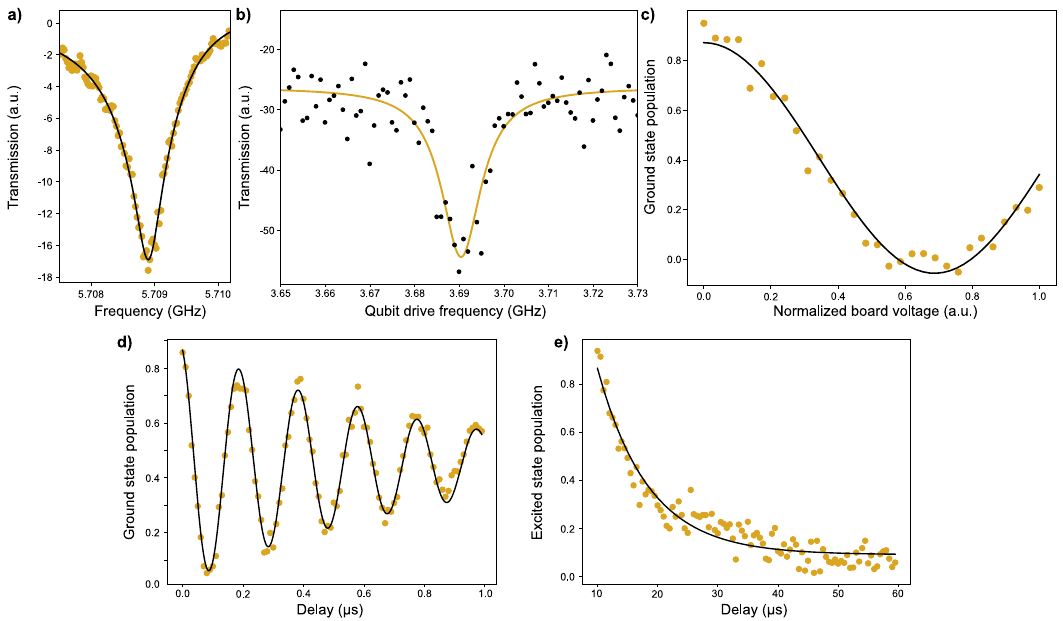}
    \caption{Summary of all-RF qubit bring-up using power domain $\pi$-pulses. We present (a) readout resonator transmission at 5.709 GHz, (b) qubit spectroscopy when the qubit is flux-biased to 3.690 GHz, (c) power Rabi oscillations reported in arbitrary voltage units, (d) Ramsey measurements, from which we extract a coherence time $T_2^*=778$ ns, and (e) a lifetime measurement, from which we extract a lifetime of $T_1=8.42$ $\upmu$s.}
    \label{fig:bringup}
\end{figure*}

\begin{figure}
    \centering
    \includegraphics{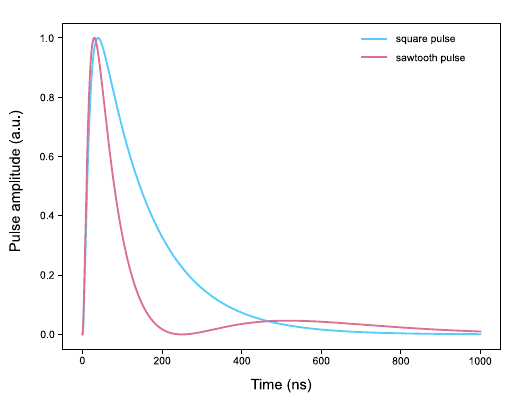}
    \caption{Simulated microwave emission profiles for 100 ns pulses. Pulse profiles are for (a) square-wave input optical fields and (b) sawtooth-wave input optical fields. }
    \label{fig:pulse_shapes}
\end{figure}

\begin{figure*}[!ht]
    \centering
    \includegraphics{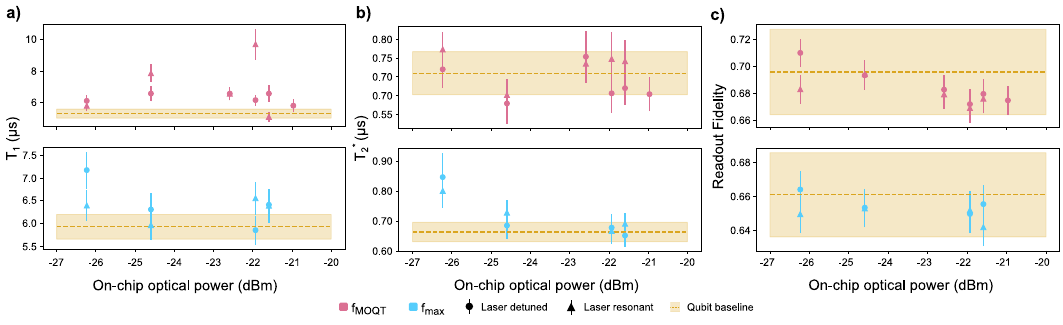}
    \caption{Qubit performance while transducer is optically pumped. We characterize (a)  Qubit lifetime ($T_1$),  (b) coherence time ($T_2^*$) , and (c) readout fidelity as a function of on-chip optical power with a single strong pump field (no optical idler) averaged over 5000 measurements. Errorbars correspond to the standard error over all repetitions. The top (pink) is measured with the qubit flux-biased to the transducer frequency ($f_{MOQT}$) while the bottom (blue) is measured with the qubit flux-biased to the qubit maximum frequency ($f_{max}$).  We report the $T_1$, $T_2^*$, and readout fidelity of the qubit while the laser is off as a dashed gold line, with the standard error shaded in gold.  We measure the qubit while the laser is off-resonant with our optical modes (circles) and locked to the red optical resonance (triangles). We see that the qubit lifetimes (a), coherence times (b), and fidelities (c) do not experience significant degradation with increased optical power at the transducer flux bias ($f_{CEO-MOQT}=3.71$ GHz) or the qubit maximum frequency ($f_{max}$=4.571 GHz) within measurement error. }
    \label{fig:decoh}
\end{figure*}


\end{document}